\documentclass[12pt]{article} 
\usepackage[T1]{fontenc}
\usepackage[utf8]{inputenc} 

\usepackage[square,comma,sort&compress,numbers]{natbib}
\usepackage{enumerate}
\usepackage{amsmath}
\usepackage{hyperref}
\usepackage{color}
\usepackage{tikz}
\usepackage{subcaption}

\newtheorem{myexample}{Example}

\definecolor{MyGrey}{RGB}{100,120,100}
\definecolor{MyYellow}{RGB}{218,165,32}
\definecolor{MyGreen}{RGB}{0,102,51}
\definecolor{MyPurple}{RGB}{102,10,153}

\usetikzlibrary{arrows.meta,
                positioning,
                shapes}
\usepackage{tabularx}
\usepackage{xurl,adjustbox, realboxes, rotating, dsfont, etoolbox, nccmath, paracol, multirow,  setspace,  fancyhdr, booktabs}
\usepackage{xcolor, graphicx, array, caption, colortbl}
\usepackage{float}
\usepackage{titlesec}
\usepackage[normalem]{ulem}
\usepackage{anyfontsize}
\DeclareMathDelimiter{(}{\mathopen} {operators}{"28}{largesymbols}{"00}
\DeclareMathDelimiter{)}{\mathclose}{operators}{"29}{largesymbols}{"01}
\usepackage[colorinlistoftodos]{todonotes}
\begin{document}
\title{A Survey on Optimization Studies of Group Centrality Metrics}

\author{Mustafa Can Camur \thanks{General Electric  Research, 1 Research Circle, Niskayuna, NY 12309} \\    \and Chrysafis Vogiatzis \thanks{
University of Illinois Urbana-Champaign, 104 S Matthews Ave, Urbana, IL 61801}}
\maketitle

\begin{abstract}

Centrality metrics have become a popular concept in network science and optimization. Over the years, centrality has been used to assign importance and  identify influential elements in various settings, including transportation, infrastructure, biological, and social networks, among others. That said, most of the literature has focused on nodal versions of centrality. Recently, group counterparts of centrality have started attracting scientific and practitioner interest. The identification of sets of nodes that are influential within a network is becoming increasingly more important. This is even more pronounced when these sets of nodes are required to induce a certain motif or structure. In this study, we review group centrality metrics from an operations research and optimization perspective for the first time. This is particularly interesting due to the rapid evolution and development of this area in the operations research community over the last decade.  We first present a historical overview of how we have reached this point in the study of group centrality. We then discuss the different structures and motifs that appear prominently in the literature, alongside the techniques and methodologies that are popular. We finally present possible avenues and directions for future work, mainly in three areas: (i) probabilistic metrics to account for randomness along with stochastic optimization techniques; (ii) structures and relaxations that have not been yet studied; and (iii) new emerging applications that can take advantage of group centrality. Our survey offers a concise review of group centrality and its intersection with network analysis and optimization.

\end{abstract}






\section{Introduction}

Centrality is {one of the most extensively} studied {concepts} in network science, graph theory, and  operations research (OR) communities. It primarily focuses on the significance, influence, and/or criticality of elements in a given network consisting of nodes and edges. Centrality metrics are inherently fundamental in comprehending complex systems from a network topology perspective. This is immediately visible from the large number of studies and applications that employ them.

Applications of centrality are wide-reaching and highly interdisciplinary. As an example, from a supply chain perspective, centrality can indicate the importance of a particular supplier (or group of suppliers) to the success of  supply chain operations. It can help decision makers identify potential bottlenecks in the supply chain network \citep{borgatti2009social}. Similarly, from an infrastructure network perspective, centrality might express the importance of connection points in ensuring a reliable and resilient network in the face of disruptive and cascading events \citep{stergiopoulos2015risk}. In fact, choosing the appropriate centrality measure  is a crucial step in every application. We refer the reader to see {Chebotarev and Gubanov} \cite{chebotarev2020choose} for an extensive discussion on how to choose the right centrality metric depending on the application.

Node-based centrality metrics (nodal centrality) are indices that aim to quantify the importance of a single node in the grand scheme of a network. In the beginning, the notion of point centrality, a precursor to the more famous nowadays betweenness centrality, was introduced in the fundamental work by Bavelas \cite{bavelas1948mathematical}. Motivated and inspired by this work, more and more researchers from a wide variety of fields started focusing on other forms of centrality, including but not limited to degree \citep{brodka2011degree}, closeness \citep{okamoto2008ranking}, betweenness \citep{freeman1977set,freeman1978centrality,riondato2014fast} and its more modern adaptation to flow betweenness \citep{freeman1991centrality}, and eigenvector centrality \citep{bonacich1972factoring,maharani2014degree} have been widely studied, applied, and taught. In fact, several survey and review studies have focused on node-based centrality metrics in the literature, especially in the context of network resiliency \citep{wan2021survey}, biological networks \citep{wang2022mini}, wireless sensor networks \citep{jain2013node}, and general complex networks \citep{saxena2020centrality}, among others.

Group centrality metrics, on the other hand, focus not on individual nodes but rather on identifying clusters or groups of network elements that collectively maximize or minimize specific criteria \citep{everett1999centrality}.  {They play a pivotal role as they help identify influential groups of nodes, which is more applicable in scenarios where collective influence or group structure is crucial. For example, in social networks, a group of interconnected individuals may have a greater combined influence on information spread than the same individuals considered separately. In infrastructure networks, such as transportation or utility networks, group centrality can identify critical clusters of components whose simultaneous failure could lead to significant disruptions. This perspective is not captured by node centrality, which only assesses the importance of individual components.}

From an application standpoint, researchers  have applied group centrality metrics to various domains, including social networks \citep{das2018study}, infrastructure networks \citep{michalak2013efficient}, biological networks \citep{vogiatzis2019identification}, traffic networks \citep{ohara2017maximizing}, human disease networks \citep{muhiuddin2023study}, and transportation networks \citep{vogiatzis2016evacuation}. Overall, group centrality has become a vital tool in the analysis of networks, one that provides a unique perspective on network dynamics in various fields by analyzing the collective behavior of its interconnected elements. 

It is important to mention that there is  growing interest in the OR community regarding primarily cohesive group \citep{frank1995identifying} centrality metrics and their applications, especially in the last decade. Given a network, a cohesive group may represent the group of nodes which are more densely interconnected among themselves than with nodes outside the group. This, intentionally open-ended, definition allows for different interpretations in different contexts.

{We note that group centrality is examined as a sub category under centrality measures in different  survey papers (see Saxena and Iyengar~\cite{saxena2020centrality} and  Wan et al.~\cite{wan2021survey}.)}
However, to the best of our knowledge, no review study has been conducted on specifically group centrality metrics from an optimization {and OR} perspective.  Hence, our work intends to provide a concise  survey of how group centrality metrics have been evolved over time and how they have been employed by OR researchers to develop new optimization models and methodologies.


 In our work, we are primarily interested in studies that aim to identify the best sets of nodes forming groups with specific objectives and use optimization methods for this identification. In large-scale networks, there could potentially be thousands, if not tens of thousands, of such subgroups. Hence, specialized optimization techniques including mathematical modeling and advanced solutions approaches (such as combinatorial branch-and-bound or Benders decomposition) play a crucial role in solving this challenging task in real-life, large-scale networked systems. 











Our work is outlined as follows. We begin with presenting preliminaries, definitions and extensions regarding both node and group centrality in Section \ref{Definitions}. Subsequently, we provide a history behind group centrality metrics from the optimization perspective (see Section \ref{History}).   In the following section, the solution approaches and applications are briefly discussed (see Section \ref{Solution Approaches}). In Section \ref{Future Work}, we present the future research direction and important gaps in the field. Lastly, we summarize our work in Section \ref{Conclusion}.

\section{Definitions and extensions} \label{Definitions}


In this section, we provide details regarding the concepts used throughout the paper. 
Let us assume that we are given a simple, undirected, and connected (i.e., there exists a path between each part of nodes) network $G =\left(V,E\right)$. We refer the reader to {Freeman} \citep{freeman1977set} for detailed discussion on centrality in disconnected graphs. As is common, $V = \{1,2, \cdots, n\}$ and $E \subseteq V \times V$ represent the set of nodes and edges, respectively. Given a node $v \in V$, let $N(v)$ be the set of nodes that are neighbors to $v$: mathematically, $N(v)=\left\{u: (u,v)\in E \right\}$. 

Moreover, let $\mathcal{P}_{ut}$ be the set of all shortest (geodesic) paths connecting nodes $u$ and $t$ ($\mathcal{P}_{ut}\neq \emptyset$, due to connectedness).  We use a similar notation for the number of shortest paths connecting $u$ to $t$ that pass through node $v\neq u, t$. Specifically, we let this quantity be $\sigma_{ut}(v)$. For simplicity, we let $\sigma_{ut}(\cdot)=\left|\mathcal{P}_{ut}\right|$.

Before proceeding to our definitions, we will use $d_{uv}$ be the length of the shortest path connecting nodes $u$ and $v$. Note that by definition, and path in $\mathcal{P}_{uv}$ will have length $d_{uv}$. We are now ready to present the definitions of three fundamental centrality metrics. For a given node $v$, we define:

\begin{itemize}
    \item \uline{Degree centrality}: The number of adjacent nodes to $v$ represented as $\left|N\left(v\right)\right|$. Often, we report the normalized version of that quantity as $\frac{\left|N\left(v\right)\right|}{\left|V\right|-1}$.
    \item \uline{Betweenness centrality}: The fraction of the number of shortest paths going through node $v$ for a any two nodes $u,t \in V \setminus \{ v\}$. It is calculated as $\sum_{u,t \in V}^{} \frac{\sigma_{ut}(v)}{\sigma_{ut}(\cdot)}$.
    \item \uline{Closeness centrality}: The inverse of the sum of shortest path between $v$ and all other nodes and is denoted as $\sum_{u \in V}^{} \frac{1}{d_{uv}}$. {We note here that some works calculate closeness centrality as $\frac{1}{\sum_{u \in V}^{} d_{uv}}$.}
\end{itemize}

{The three above centrality metrics \cite{freeman2002centrality} have been traditionally used as the basis of development for other variants of centrality. Specifically, betweenness and closeness are both based on shortest paths in a network. However, as noted in \cite{stephenson1989rethinking}, using shortest paths can sometimes be limiting. Calculating then betweenness and closeness using all paths (or all paths bounded by some distance threshold) leads to the development of information or current-flow betweenness and closeness centrality metrics.}

Given a subset of nodes $S \subseteq V$, we define $G\left[S\right]$ as the induced subgraph where the edge set is $S \times S \cap E$. In this work, we use $G\left[S\right]$ for any group structure. Then, the centrality concepts introduced above can be extended to sets of nodes $S\subseteq V$ with a slight misuse of notation.

\begin{itemize}
    \item \uline{Group degree centrality}: Formally, we define the degree centrality of a set of nodes $S$ as the number of distinct adjacent nodes to any node in $S$. We let $N(S)$ be the open neighborhood of node set $S$: $N(S)=\left\{u: (u,v)\in E, \forall v\in S \right\}$. Then, group degree centrality is calculated as $|N(S)|$ (or $\frac{|N(S)|}{|V|-|S|}$ in the normalized version).
    \item \uline{Group betweenness centrality}: The ratio of  shortest paths going through any node in $S$ for  a any two nodes  $u,t \in V \setminus \{ S\}$ denoted as $\sum_{u,t \in V}^{} \frac{\sigma_{ut}(S)}{\sigma_{ut}(.)}$.
    \item \uline{Group closeness centrality}: The inverse of the sum of shortest path between any node in $S$ and all other nodes and is denoted as $\sum_{u \in V}^{} \frac{1}{d_{uS}}$, where $d_{uS}=\min_{v\in S} d_{uv}$ (i.e., $d_{uS}$ is the shortest path distance from node $u$ to any node in $S$). We note that this quantity is also known as group eccentricity.
\end{itemize}

\begin{figure}
\centering
    \resizebox{\textwidth}{!}{\begin{tikzpicture}[transform shape, thick, minimum size=4mm, inner sep=0pt]
	\node[circle, draw, fill=blue!20] (N-1) at (0,0) {\tiny 1};
	\node[circle, draw, fill=red!40, right of=N-1] (N-2) {\tiny 2};
    \node[circle, draw, fill=red!40, below of=N-2] (N-4) {\tiny 4};
    \node[circle, draw, fill=blue!20, above of=N-2] (N-5) {\tiny 5};
    \node[circle, draw, fill=blue!20, below of=N-4] (N-7) {\tiny 7};
	\node[circle, draw, fill=blue!20, right of=N-2] (N-3) {\tiny 3};
    \node[circle, draw, fill=red!40, above of=N-3] (N-6) {\tiny 6};
    \node[circle, draw, fill=blue!20, right of=N-3] (N-8) {\tiny 8};
    \node[circle, draw, fill=red!40, right of=N-8] (N-9) {\tiny 9};
    \node[circle, draw, fill=blue!20, below of=N-8] (N-10) {\tiny 10};
    \node[circle, draw, fill=blue!20, above of=N-8] (N-12) {\tiny 12};
    \node[circle, draw, fill=blue!20, below of=N-9] (N-11) {\tiny 11};
    \node[circle, draw, fill=blue!20, right of=N-9] (N-13) {\tiny 13};
    \node[circle, draw, fill=blue!20, right of=N-13] (N-15) {\tiny 15};
    \node[circle, draw, fill=blue!20, below of=N-15] (N-16) {\tiny 16};
    \node[circle, draw, fill=blue!20, above of=N-15] (N-14) {\tiny 14};

	\path (N-1) edge[-, thick] (N-2);
    \path (N-2) edge[-, thick] (N-3);
    \path (N-2) edge[-, thick] (N-4);
    \path (N-2) edge[-, thick] (N-5);
    \path (N-3) edge[-, thick] (N-4);
    \path (N-3) edge[-, thick] (N-6);
    \path (N-3) edge[-, thick] (N-8);
    \path (N-4) edge[-, thick] (N-7);
    \path (N-5) edge[-, thick] (N-6);
    \path (N-6) edge[-, thick] (N-12);
    \path (N-8) edge[-, thick] (N-9);
    \path (N-8) edge[-, thick] (N-10);
    \path (N-8) edge[-, thick] (N-11);
    \path (N-8) edge[-, thick] (N-12);

    \path (N-9) edge[-, thick] (N-13);
    \path (N-9) edge[-, thick] (N-10);
    \path (N-9) edge[-, thick] (N-11);
    \path (N-10) edge[-, thick] (N-11);
    \path (N-13) edge[-, thick] (N-16);
    \path (N-13) edge[-, thick] (N-14);
    \path (N-13) edge[-, thick] (N-15);

\end{tikzpicture}}
    \caption{The network for Example \ref{ex:group1}. The set of nodes selected to be in the group (nodes $S=\left\{2,4,6,9\right\}$) do not induce a connected subgraph.  \label{group_centrality_example}}
\end{figure}
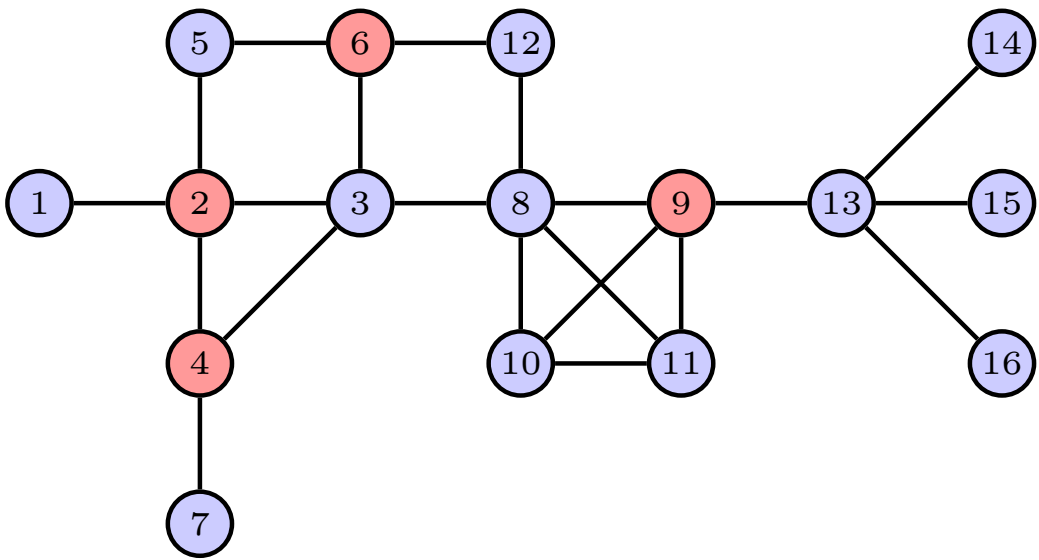

{In some instances (like in closeness centrality), it may make sense to define a bottleneck version (or, equivalently, maximum distance closeness centrality). In that version, the biggest difference is that we are focused on the worst-case possible distance from any node outside $S$ to some node in $S$. In mathematical terms, the maximum distance closeness centrality is defined as $1/\max_{u\in V} d_{uS}$, where $d_{uS}=\min_{v\in S} d_{uv}$.} We present Example \ref{ex:group1} (based on Figure~\ref{group_centrality_example}) showing the calculation for the three definitions above for a set of nodes that does not induce a connected subgraph.

\begin{myexample}
    \label{ex:group1}

    In Figure \ref{group_centrality_example}, we have selected our group to be $S=\left\{2,4,6,9\right\}$. Then, the group degree centrality is $|N(S)|=9$ or $\frac{|N(S)|}{|V\setminus S|}=9/12=0.75$ in the normalized version.
    
    The group closeness centrality is $9\cdot \frac{1}{1}+3\cdot \frac{1}{2}=\frac{21}{2}$, seeing as $9$ nodes are at a distance of $1$ and $3$ nodes are at a distance of $2$ from at least one node in $S$. 

    Finally, the group betweenness centrality is calculated as follows. {First, we will consider nodes in lexicographic order, for simplicity. That is, starting from node $1$, we observe that there is exactly one shortest path towards nodes $3$, $5$, $7$, $8$, $10$, $11$, $13$, $14$, $15$, and $16$ and it always uses at least one node in $S$; furthermore, it is only node $12$ that has three shortest paths originating from $1$, however all three again use at least one node in $S$. This brings us to a total of 11. Continuing to node $3$, we have 2 shortest paths towards $5$ with both using a node in $S$ and 1 shortest path towards $7$ that also uses a node in $S$. Similarly, we have only one shortest path towards nodes $8, 10, 11$ which are not passing through a node in $S$, as well as one shortest path towards nodes $13, 14, 15, 16$, all of which are passing through a node in $S$. We left the shortest paths from node $3$ to node $12$ for last: there are two shortest paths, only one of which is passing through a node in $S$ (namely, node $6$). Summing up all the fractions, we have a total of $7.5$.}
    
    {Moving on, we can calculate that same count originating from nodes $5$ (9, when excluding shortest paths towards $1$ and $3$) and $7$ (8, excluding shortest paths towards $1, 3, 5$). Nodes $8, 10, 11, 12$ have summations equal to $4$. This brings our total to $11+6.5+9+8+4+4+4+4=50.5$.}


    {All of the calculations for betweenness thus far are not normalized. A normalized value can be computed by dividing the obtained score by $\frac{(|V|-|S|)\cdot (|V|-|S|-1)}{2}$. In our previous calculation, where $|V|=16$ and $|S|=4$, the normalized calculation would lead to $50.5/66=0.765.$}
\end{myexample}

When set $S$ is required to induce a specific ``motif'' or structure (such as a clique or a star), then we are dealing with the centrality of induced clusters. We provide here a definition of some of the most prominently used types of structures here.

To begin with, a \textit{walk} is a list of edges such that each edge has exactly one node in common with the previous one. Note how this definition allows for a repetition of nodes (e.g., $\left\{\left(u_1, u_2\right), \left(u_2, u_3\right),\left(u_3, u_1\right), \left(u_1, u_4\right), \ldots\right\}$ is a walk as node $u_1$ is visited twice). A \textit{path} is then a walk that disallows any repetitions of nodes. A path where the starting and ending nodes are the same is a \textit{cycle}. Walks, paths, and cycles induce a connected subgraph, by definition. A connected subgraph allows for any two nodes in the subgraph to be connected using a path that is all within the subgraph. The largest connected subgraph in a graph is called a \textit{component}. A connected graph will only have one component: the full graph itself.

A set of nodes $C\subseteq V$ forms a clique if $C$ is inducing a complete subgraph \citep{pardalos1994maximum}. In other words, $C\subseteq V$ forms a clique if all edges $(i,j), i\in C, j\in C$ are present ($E[G[C]]=C\times C$). On the other hand, $C\subseteq V$ is said to form a star if there exists exactly one node $s\in C$ such that $\left(s,j\right)\in E$ for all other nodes $j\in C\setminus\left\{s\right\}$ and no other edge $\left(i,j\right)$ exists between any two nodes $i,j\in C\setminus\left\{s\right\}$. We observe how $s$ has a prominent role and is referred to as the center of the star, whereas all other nodes $i\in C\setminus\left\{s\right\}$ are referred to as the leaves of the star. We also note that both cliques and stars appear to have a relationship to independent sets, where any two nodes cannot be adjacent. A clique in graph $G\left(V,E\right)$ is an independent set in the complement graph $G^\prime\left(V,E^\prime\right)$ where $\left(i,j\right)\in E^\prime$ if and only if $\left(i,j\right)\notin E$. The leaves in the star in $G\left(V,E\right)$ also induce an independent set {within} $G$. 

Cliques and stars are both ideal situations. It may be impractical to expect that either all edges exist between any two nodes in the whole group (clique) or that no edges exist between most of the nodes in the group (star). For a star, the easier relaxation would be to allow any two nodes to be adjacent, so long as they both have an edge connecting them to the center. This is referred to as a \textit{representative}. For cliques, the situation is more complex due to the number of properties that can be relaxed. We discuss some of these relaxations next. 

The diameter is the length/cost of the longest of all shortest paths in the network. In a clique, the diameter is equal to $1$ as all shortest paths are between immediately adjacent nodes. Relaxing the diameter restriction affects the group's reachability. A group of nodes inducing a subgraph with diameter smaller than or equal to $k$ is referred to as a \textit{$k$-club} \citep{mokken1979cliques}. The density of a graph is the fraction of edges that are present compared to all possible edges being present. The clique, being an ideal subgraph, has density equal to 1 as all possible edges are present. Relaxing this restriction allows some of the edges to be absent and affects the group's familiarity. If fraction $\gamma\in\left[0,1\right]$ of all possible edges is present, then this group forms a \textit{$\gamma$-quasi-clique} \citep{abello2002massive}. As is easy to check $\gamma=1$ is the same as a clique, and $\gamma=0$ can be any group of nodes. 

Next, we focus on degree (as defined earlier): clique members $i\in C$ always have a degree of $|C|-1$; relaxing this requirement to having a degree of at least $|C|-k$ leads to the definition of \textit{$k$-plex} \citep{seidman1978graph,balasundaram2011clique}. We can also see another common clique relaxation for the members' degrees: a \textit{$k$-core} is defined as the induced subgraph where all nodes have minimum degree equal to $k$ \citep{seidman1983network}. Last, we look at connectivity (or the group's robustness). To disconnect a clique, we would need to remove almost all of its members. Specifically, to produce a disconnected graph after removing members of a clique $C$, we would need to delete $|C|-1$ of them. On the other hand, relaxing this to disconnect after removal of $|C|-k$ nodes results in the definition of a \textit{$k$-block}. 

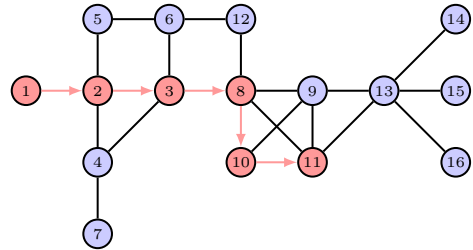
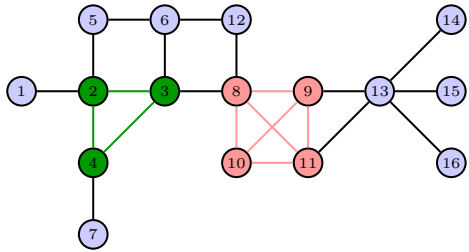
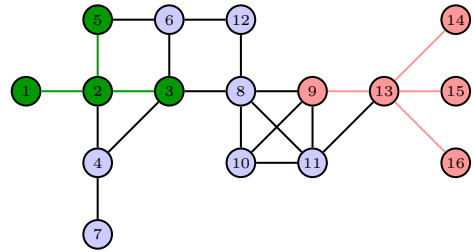
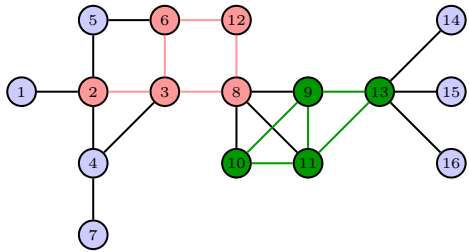
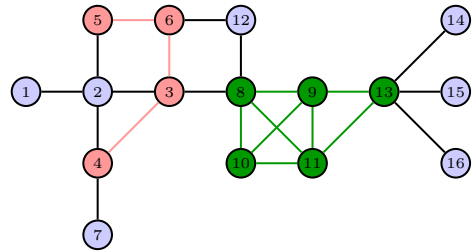
\begin{figure}
\begin{subfigure}[t]{0.47\textwidth}
        \centering
        \resizebox{.95\textwidth}{!}{\begin{tikzpicture}[transform shape, thick, minimum size=4mm, inner sep=0pt]
	\node[circle, draw, fill=blue!20] (N-1) at (0,0) {\tiny 1};
	\node[circle, draw, fill=red!40, right of=N-1] (N-2) {\tiny 2};
    \node[circle, draw, fill=blue!20, below of=N-2] (N-4) {\tiny 4};
    \node[circle, draw, fill=blue!20, above of=N-2] (N-5) {\tiny 5};
    \node[circle, draw, fill=blue!20, below of=N-4] (N-7) {\tiny 7};
	\node[circle, draw, fill=red!40, right of=N-2] (N-3) {\tiny 3};
    \node[circle, draw, fill=blue!20, above of=N-3] (N-6) {\tiny 6};
    \node[circle, draw, fill=red!40, right of=N-3] (N-8) {\tiny 8};
    \node[circle, draw, fill=red!40, right of=N-8] (N-9) {\tiny 9};
    \node[circle, draw, fill=red!40, below of=N-8] (N-10) {\tiny 10};
    \node[circle, draw, fill=blue!20, above of=N-8] (N-12) {\tiny 12};
    \node[circle, draw, fill=red!40, below of=N-9] (N-11) {\tiny 11};
    \node[circle, draw, fill=blue!20, right of=N-9] (N-13) {\tiny 13};
    \node[circle, draw, fill=blue!20, right of=N-13] (N-15) {\tiny 15};
    \node[circle, draw, fill=blue!20, below of=N-15] (N-16) {\tiny 16};
    \node[circle, draw, fill=blue!20, above of=N-15] (N-14) {\tiny 14};

	\path (N-1) edge[-, thick] (N-2);
    \path (N-3) edge[-latex, thick, red!40, bend right=30] (N-2);
    \path (N-2) edge[-, thick] (N-3);
    \path (N-2) edge[-, thick] (N-4);
    \path (N-2) edge[-, thick] (N-5);
    \path (N-3) edge[-, thick] (N-4);
    \path (N-3) edge[-, thick] (N-6);
    \path (N-3) edge[-, thick] (N-8);
    \path (N-8) edge[-latex, thick, red!40, bend right=30] (N-3);
    \path (N-4) edge[-, thick] (N-7);
    \path (N-5) edge[-, thick] (N-6);
    \path (N-6) edge[-, thick] (N-12);
    \path (N-8) edge[-, thick] (N-9);
    \path (N-8) edge[-latex, thick, red!40, bend left=30] (N-9);
    \path (N-8) edge[-, thick] (N-10);
    \path (N-10) edge[-latex, thick, bend left=30, red!40] (N-8);
    \path (N-8) edge[-, thick] (N-11);
    \path (N-8) edge[-, thick] (N-12);

    \path (N-9) edge[-, thick] (N-13);
    \path (N-9) edge[-, thick] (N-10);
    \path (N-9) edge[-, thick] (N-11);
    \path (N-9) edge[-latex, thick, red!40, bend left=30] (N-11);
    \path (N-11) edge[-latex, thick, bend left=30, red!40] (N-10);
    \path (N-10) edge[-, thick] (N-11);
    \path (N-11) edge[-, thick] (N-13);
    \path (N-13) edge[-, thick] (N-16);
    \path (N-13) edge[-, thick] (N-14);
    \path (N-13) edge[-, thick] (N-15);

\end{tikzpicture}}
        \caption{Walk $8\rightarrow 9\rightarrow 10\rightarrow 11\rightarrow 8\rightarrow 3\rightarrow 2$. If the walk ended when we visited node $8$ for the second time then it would form a cycle.}
    \end{subfigure}\hfill
\begin{subfigure}[t]{0.47\textwidth}
        \centering
        \resizebox{.95\textwidth}{!}{\begin{tikzpicture}[transform shape, thick, minimum size=4mm, inner sep=0pt]
	\node[circle, draw, fill=red!40] (N-1) at (0,0) {\tiny 1};
	\node[circle, draw, fill=red!40, right of=N-1] (N-2) {\tiny 2};
    \node[circle, draw, fill=blue!20, below of=N-2] (N-4) {\tiny 4};
    \node[circle, draw, fill=blue!20, above of=N-2] (N-5) {\tiny 5};
    \node[circle, draw, fill=blue!20, below of=N-4] (N-7) {\tiny 7};
	\node[circle, draw, fill=red!40, right of=N-2] (N-3) {\tiny 3};
    \node[circle, draw, fill=blue!20, above of=N-3] (N-6) {\tiny 6};
    \node[circle, draw, fill=red!40, right of=N-3] (N-8) {\tiny 8};
    \node[circle, draw, fill=blue!20, right of=N-8] (N-9) {\tiny 9};
    \node[circle, draw, fill=red!40, below of=N-8] (N-10) {\tiny 10};
    \node[circle, draw, fill=blue!20, above of=N-8] (N-12) {\tiny 12};
    \node[circle, draw, fill=red!40, below of=N-9] (N-11) {\tiny 11};
    \node[circle, draw, fill=blue!20, right of=N-9] (N-13) {\tiny 13};
    \node[circle, draw, fill=blue!20, right of=N-13] (N-15) {\tiny 15};
    \node[circle, draw, fill=blue!20, below of=N-15] (N-16) {\tiny 16};
    \node[circle, draw, fill=blue!20, above of=N-15] (N-14) {\tiny 14};

	\path (N-1) edge[-latex, thick, red!40] (N-2);
    \path (N-2) edge[-latex, thick, red!40] (N-3);
    \path (N-2) edge[-, thick] (N-4);
    \path (N-2) edge[-, thick] (N-5);
    \path (N-3) edge[-, thick] (N-4);
    \path (N-3) edge[-, thick] (N-6);
    \path (N-3) edge[-latex, thick, red!40] (N-8);
    \path (N-4) edge[-, thick] (N-7);
    \path (N-5) edge[-, thick] (N-6);
    \path (N-6) edge[-, thick] (N-12);
    
    \path (N-8) edge[-, thick] (N-9);
    \path (N-8) edge[-latex, thick, red!40] (N-10);
    \path (N-8) edge[-, thick] (N-11);
    \path (N-8) edge[-, thick] (N-12);

    \path (N-9) edge[-, thick] (N-13);
    \path (N-9) edge[-, thick] (N-10);
    \path (N-9) edge[-, thick] (N-11);
    \path (N-10) edge[-latex, thick, red!40] (N-11);
    \path (N-11) edge[-, thick] (N-13);
    \path (N-13) edge[-, thick] (N-16);
    \path (N-13) edge[-, thick] (N-14);
    \path (N-13) edge[-, thick] (N-15);

\end{tikzpicture}}
        \caption{Path $1\rightarrow 2\rightarrow 3\rightarrow 8\rightarrow 10\rightarrow 11$. }
    \end{subfigure}

    \begin{subfigure}[t]{0.47\textwidth}
        \centering
        \resizebox{.95\textwidth}{!}{\begin{tikzpicture}[transform shape, thick, minimum size=4mm, inner sep=0pt]
	\node[circle, draw, fill=blue!20] (N-1) at (0,0) {\tiny 1};
	\node[circle, draw, fill=green!60!black, right of=N-1] (N-2) {\tiny 2};
    \node[circle, draw, fill=green!60!black, below of=N-2] (N-4) {\tiny 4};
    \node[circle, draw, fill=blue!20, above of=N-2] (N-5) {\tiny 5};
    \node[circle, draw, fill=blue!20, below of=N-4] (N-7) {\tiny 7};
	\node[circle, draw, fill=green!60!black, right of=N-2] (N-3) {\tiny 3};
    \node[circle, draw, fill=blue!20, above of=N-3] (N-6) {\tiny 6};
    \node[circle, draw, fill=red!40, right of=N-3] (N-8) {\tiny 8};
    \node[circle, draw, fill=red!40, right of=N-8] (N-9) {\tiny 9};
    \node[circle, draw, fill=red!40, below of=N-8] (N-10) {\tiny 10};
    \node[circle, draw, fill=blue!20, above of=N-8] (N-12) {\tiny 12};
    \node[circle, draw, fill=red!40, below of=N-9] (N-11) {\tiny 11};
    \node[circle, draw, fill=blue!20, right of=N-9] (N-13) {\tiny 13};
    \node[circle, draw, fill=blue!20, right of=N-13] (N-15) {\tiny 15};
    \node[circle, draw, fill=blue!20, below of=N-15] (N-16) {\tiny 16};
    \node[circle, draw, fill=blue!20, above of=N-15] (N-14) {\tiny 14};

	\path (N-1) edge[-, thick] (N-2);
    \path (N-2) edge[-, thick, green!60!black] (N-3);
    \path (N-2) edge[-, thick, green!60!black] (N-4);
    \path (N-2) edge[-, thick] (N-5);
    \path (N-3) edge[-, thick, green!60!black] (N-4);
    \path (N-3) edge[-, thick] (N-6);
    \path (N-3) edge[-, thick] (N-8);
    \path (N-4) edge[-, thick] (N-7);
    \path (N-5) edge[-, thick] (N-6);
    \path (N-8) edge[-, thick, red!40] (N-9);
    \path (N-8) edge[-, thick, red!40] (N-10);
    \path (N-8) edge[-, thick, red!40] (N-11);
    \path (N-8) edge[-, thick] (N-12);

    \path (N-9) edge[-, thick] (N-13);
    \path (N-9) edge[-, thick, red!40] (N-10);
    \path (N-9) edge[-, thick, red!40] (N-11);
    \path (N-10) edge[-, thick, red!40] (N-11);
    \path (N-11) edge[-, thick] (N-13);
    \path (N-13) edge[-, thick] (N-16);
    \path (N-13) edge[-, thick] (N-14);
    \path (N-13) edge[-, thick] (N-15);
    
    \path (N-6) edge[-, thick] (N-12);
\end{tikzpicture}}
        \caption{A clique of size $3$ (in green) and a clique of size $4$ (in red). Any subset of these two groups is, by definition, also a clique due to the hereditary property of cliques.}
    \end{subfigure}\hfill
    \begin{subfigure}[t]{0.47\textwidth}
        \centering
        \resizebox{.95\textwidth}{!}{\begin{tikzpicture}[transform shape, thick, minimum size=4mm, inner sep=0pt]
	\node[circle, draw, fill=green!60!black] (N-1) at (0,0) {\tiny 1};
	\node[circle, draw, fill=green!60!black, right of=N-1] (N-2) {\tiny 2};
    \node[circle, draw, fill=blue!20, below of=N-2] (N-4) {\tiny 4};
    \node[circle, draw, fill=green!60!black, above of=N-2] (N-5) {\tiny 5};
    \node[circle, draw, fill=blue!20, below of=N-4] (N-7) {\tiny 7};
	\node[circle, draw, fill=green!60!black, right of=N-2] (N-3) {\tiny 3};
    \node[circle, draw, fill=blue!20, above of=N-3] (N-6) {\tiny 6};
    \node[circle, draw, fill=blue!20, right of=N-3] (N-8) {\tiny 8};
    \node[circle, draw, fill=red!40, right of=N-8] (N-9) {\tiny 9};
    \node[circle, draw, fill=blue!20, below of=N-8] (N-10) {\tiny 10};
    \node[circle, draw, fill=blue!20, above of=N-8] (N-12) {\tiny 12};
    \node[circle, draw, fill=blue!20, below of=N-9] (N-11) {\tiny 11};
    \node[circle, draw, fill=red!40, right of=N-9] (N-13) {\tiny 13};
    \node[circle, draw, fill=red!40, right of=N-13] (N-15) {\tiny 15};
    \node[circle, draw, fill=red!40, below of=N-15] (N-16) {\tiny 16};
    \node[circle, draw, fill=red!40, above of=N-15] (N-14) {\tiny 14};

	\path (N-1) edge[-, thick, green!60!black] (N-2);
    \path (N-2) edge[-, thick, green!60!black] (N-3);
    \path (N-2) edge[-, thick] (N-4);
    \path (N-2) edge[-, thick, green!60!black] (N-5);
    \path (N-3) edge[-, thick] (N-4);
    \path (N-3) edge[-, thick] (N-6);
    \path (N-3) edge[-, thick] (N-8);
    \path (N-4) edge[-, thick] (N-7);
    \path (N-5) edge[-, thick] (N-6);
    \path (N-6) edge[-, thick] (N-12);
    
    \path (N-8) edge[-, thick] (N-9);
    \path (N-8) edge[-, thick] (N-10);
    \path (N-8) edge[-, thick] (N-11);
    \path (N-8) edge[-, thick] (N-12);

    \path (N-9) edge[-, thick, red!40] (N-13);
    \path (N-9) edge[-, thick] (N-10);
    \path (N-9) edge[-, thick] (N-11);
    \path (N-10) edge[-, thick] (N-11);
    \path (N-11) edge[-, thick] (N-13);
    \path (N-13) edge[-, thick, red!40] (N-16);
    \path (N-13) edge[-, thick, red!40] (N-14);
    \path (N-13) edge[-, thick, red!40] (N-15);

\end{tikzpicture}}
        \caption{A star with $3$ leaves (in green) and a star with $4$ leaves (in red). Note that we could have replaced leaf $9$ with leaf $11$. Not both $9$ and $11$ can serve as leaves though for the structure in red.}
    \end{subfigure}

    \begin{subfigure}[t]{0.47\textwidth}
        \centering
        \resizebox{.95\textwidth}{!}{\begin{tikzpicture}[transform shape, thick, minimum size=4mm, inner sep=0pt]
	\node[circle, draw, fill=blue!20] (N-1) at (0,0) {\tiny 1};
	\node[circle, draw, fill=red!40, right of=N-1] (N-2) {\tiny 2};
    \node[circle, draw, fill=blue!20, below of=N-2] (N-4) {\tiny 4};
    \node[circle, draw, fill=blue!20, above of=N-2] (N-5) {\tiny 5};
    \node[circle, draw, fill=blue!20, below of=N-4] (N-7) {\tiny 7};
	\node[circle, draw, fill=red!40, right of=N-2] (N-3) {\tiny 3};
    \node[circle, draw, fill=red!40, above of=N-3] (N-6) {\tiny 6};
    \node[circle, draw, fill=red!40, right of=N-3] (N-8) {\tiny 8};
    \node[circle, draw, fill=green!60!black, right of=N-8] (N-9) {\tiny 9};
    \node[circle, draw, fill=green!60!black, below of=N-8] (N-10) {\tiny 10};
    \node[circle, draw, fill=red!40, above of=N-8] (N-12) {\tiny 12};
    \node[circle, draw, fill=green!60!black, below of=N-9] (N-11) {\tiny 11};
    \node[circle, draw, fill=green!60!black, right of=N-9] (N-13) {\tiny 13};
    \node[circle, draw, fill=blue!20, right of=N-13] (N-15) {\tiny 15};
    \node[circle, draw, fill=blue!20, below of=N-15] (N-16) {\tiny 16};
    \node[circle, draw, fill=blue!20, above of=N-15] (N-14) {\tiny 14};

	\path (N-1) edge[-, thick] (N-2);
    \path (N-2) edge[-, thick, red!40] (N-3);
    \path (N-2) edge[-, thick] (N-4);
    \path (N-2) edge[-, thick] (N-5);
    \path (N-3) edge[-, thick] (N-4);
    \path (N-3) edge[-, thick, red!40] (N-6);
    \path (N-3) edge[-, thick, red!40] (N-8);
    \path (N-4) edge[-, thick] (N-7);
    \path (N-5) edge[-, thick] (N-6);
    \path (N-6) edge[-, thick, red!40] (N-12);
    
    \path (N-8) edge[-, thick] (N-9);
    \path (N-8) edge[-, thick] (N-10);
    \path (N-8) edge[-, thick] (N-11);
    \path (N-8) edge[-, thick, red!40] (N-12);

    \path (N-9) edge[-, thick, green!60!black] (N-13);
    \path (N-9) edge[-, thick, green!60!black] (N-10);
    \path (N-9) edge[-, thick, green!60!black] (N-11);
    \path (N-10) edge[-, thick, green!60!black] (N-11);
    \path (N-11) edge[-, thick, green!60!black] (N-13);
    \path (N-13) edge[-, thick] (N-16);
    \path (N-13) edge[-, thick] (N-14);
    \path (N-13) edge[-, thick] (N-15);

\end{tikzpicture}}
        \caption{A $\gamma$-quasi-clique for $\gamma=0.5$ in red. Note how the induced graph includes $5$ edges out of a total of $10$ possible edges. The nodes in green form a $2$-plex, as every node in the group (of cardinality $|C|=4$) has degree at least $|C|-2=2$.}
    \end{subfigure}\hfill
    \begin{subfigure}[t]{0.47\textwidth}
        \centering
        \resizebox{.95\textwidth}{!}{\begin{tikzpicture}[transform shape, thick, minimum size=4mm, inner sep=0pt]
	\node[circle, draw, fill=blue!20] (N-1) at (0,0) {\tiny 1};
	\node[circle, draw, fill=blue!20, right of=N-1] (N-2) {\tiny 2};
    \node[circle, draw, fill=red!40, below of=N-2] (N-4) {\tiny 4};
    \node[circle, draw, fill=red!40, above of=N-2] (N-5) {\tiny 5};
    \node[circle, draw, fill=blue!20, below of=N-4] (N-7) {\tiny 7};
	\node[circle, draw, fill=red!40, right of=N-2] (N-3) {\tiny 3};
    \node[circle, draw, fill=red!40, above of=N-3] (N-6) {\tiny 6};
    \node[circle, draw, fill=green!60!black, right of=N-3] (N-8) {\tiny 8};
    \node[circle, draw, fill=green!60!black, right of=N-8] (N-9) {\tiny 9};
    \node[circle, draw, fill=green!60!black, below of=N-8] (N-10) {\tiny 10};
    \node[circle, draw, fill=blue!20, above of=N-8] (N-12) {\tiny 12};
    \node[circle, draw, fill=green!60!black, below of=N-9] (N-11) {\tiny 11};
    \node[circle, draw, fill=green!60!black, right of=N-9] (N-13) {\tiny 13};
    \node[circle, draw, fill=blue!20, right of=N-13] (N-15) {\tiny 15};
    \node[circle, draw, fill=blue!20, below of=N-15] (N-16) {\tiny 16};
    \node[circle, draw, fill=blue!20, above of=N-15] (N-14) {\tiny 14};

	\path (N-1) edge[-, thick] (N-2);
    \path (N-2) edge[-, thick] (N-3);
    \path (N-2) edge[-, thick] (N-4);
    \path (N-2) edge[-, thick] (N-5);
    \path (N-3) edge[-, thick, red!40] (N-4);
    \path (N-3) edge[-, thick, red!40] (N-6);
    \path (N-3) edge[-, thick] (N-8);
    \path (N-4) edge[-, thick] (N-7);
    \path (N-5) edge[-, thick, red!40] (N-6);
    \path (N-6) edge[-, thick] (N-12);
    
    \path (N-8) edge[-, thick, green!60!black] (N-9);
    \path (N-8) edge[-, thick, green!60!black] (N-10);
    \path (N-8) edge[-, thick, green!60!black] (N-11);
    \path (N-8) edge[-, thick] (N-12);

    \path (N-9) edge[-, thick,green!60!black] (N-13);
    \path (N-9) edge[-, thick,green!60!black] (N-10);
    \path (N-9) edge[-, thick,green!60!black] (N-11);
    \path (N-10) edge[-, thick,green!60!black] (N-11);
    \path (N-11) edge[-, thick, green!60!black] (N-13);
    \path (N-13) edge[-, thick] (N-16);
    \path (N-13) edge[-, thick] (N-14);
    \path (N-13) edge[-, thick] (N-15);

\end{tikzpicture}}
        \caption{A $2$-club is shown in green. On the other hand, the nodes in red do not induce a $2$-club. While nodes $4$ and $5$ are reachable within $2$ hops, the path uses a node outside the group. The structure in green is also a $2$-core (as all nodes have degree at least 2).}
    \end{subfigure}
    \caption{A few examples from notable structures that have been proposed.}\label{fig:structures}
\end{figure}

We present examples of some of these structures in Figure \ref{fig:structures}. We also would like to note that not all of the above structures have been studied from a group centrality perspective. We present the structures that have been studied later in Table \ref{Table1} of Section \ref{Solution Approaches}.

\begin{figure}
\centering
    \resizebox{\textwidth}{!}{\begin{tikzpicture}[transform shape, thick, minimum size=4mm, inner sep=0pt]
	\node[circle, draw, fill=blue!20] (N-1) at (0,0) {\tiny 1};
	\node[circle, draw, fill=green!60!black, right of=N-1] (N-2) {\tiny 2};
    \node[circle, draw, fill=green!60!black, below of=N-2] (N-4) {\tiny 4};
    \node[circle, draw, fill=blue!20, above of=N-2] (N-5) {\tiny 5};
    \node[circle, draw, fill=blue!20, below of=N-4] (N-7) {\tiny 7};
	\node[circle, draw, fill=green!60!black, right of=N-2] (N-3) {\tiny 3};
    \node[circle, draw, fill=blue!20, above of=N-3] (N-6) {\tiny 6};
    \node[circle, draw, fill=red!40, right of=N-3] (N-8) {\tiny 8};
    \node[circle, draw, fill=red!40, right of=N-8] (N-9) {\tiny 9};
    \node[circle, draw, fill=red!40, below of=N-8] (N-10) {\tiny 10};
    \node[circle, draw, fill=blue!20, above of=N-8] (N-12) {\tiny 12};
    \node[circle, draw, fill=red!40, below of=N-9] (N-11) {\tiny 11};
    \node[circle, draw, fill=blue!20, right of=N-9] (N-13) {\tiny 13};
    \node[circle, draw, fill=blue!20, right of=N-13] (N-15) {\tiny 15};
    \node[circle, draw, fill=blue!20, above of=N-15] (N-14) {\tiny 14};

	\path (N-1) edge[-, thick] (N-2);
    \path (N-2) edge[-, thick, green!60!black] (N-3);
    \path (N-2) edge[-, thick, green!60!black] (N-4);
    \path (N-2) edge[-, thick] (N-5);
    \path (N-3) edge[-, thick, green!60!black] (N-4);
    \path (N-3) edge[-, thick] (N-6);
    \path (N-3) edge[-, thick] (N-8);
    \path (N-4) edge[-, thick] (N-7);
    
    \path (N-8) edge[-, thick, red!40] (N-9);
    \path (N-8) edge[-, thick, red!40] (N-10);
    \path (N-8) edge[-, thick, red!40] (N-11);
    \path (N-8) edge[-, thick] (N-12);

    \path (N-9) edge[-, thick] (N-13);
    \path (N-9) edge[-, thick, red!40] (N-10);
    \path (N-9) edge[-, thick, red!40] (N-11);
    \path (N-10) edge[-, thick, red!40] (N-11);
    
    \path (N-13) edge[-, thick] (N-14);
    \path (N-13) edge[-, thick] (N-15);

\end{tikzpicture}}
    \caption{The network for Examples \ref{ex:clique1} and \ref{ex:clique2}. Observe how it is slightly different to the network used in Figure \ref{fig:structures} (has one fewer node and is missing some of the previous edges).}  \label{clique_centrality_example}
\end{figure}
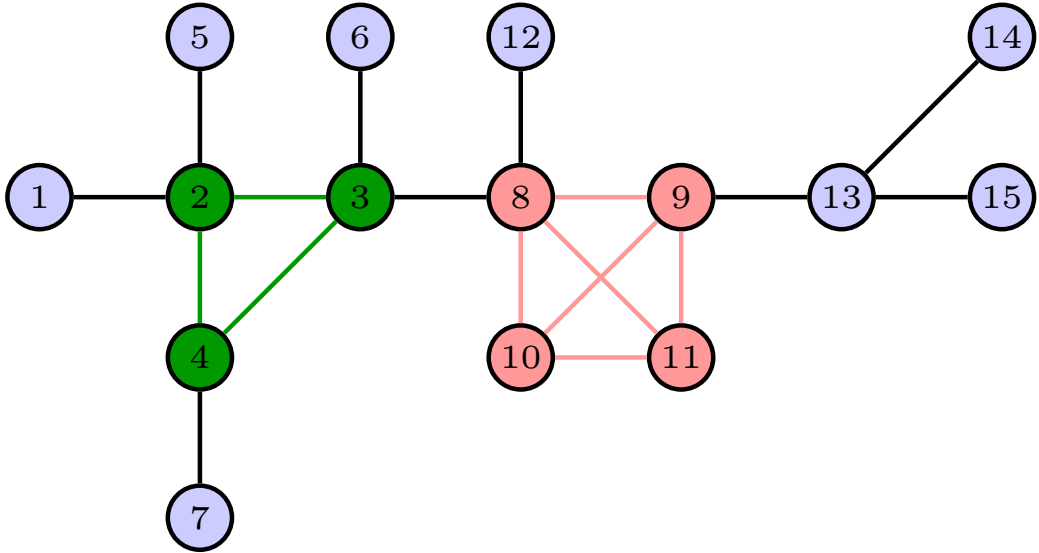

This topic has been quite prominent in the optimization community, as we will discuss in the next Section. An example of how the previously defined three notions of group centrality change in the case of cliques is discussed next based on Figure \ref{clique_centrality_example} adapted from the works in {Vogiatzis et al.} \citep{vogiatzis2015integer} and {Rasti and Vogiatzis} \citep{rasti2022novel}.

\begin{myexample}
\label{ex:clique1}
    In Figure \ref{clique_centrality_example}, we show two cliques: one of cardinality/size $3$ containing nodes $S_1=\left\{2,3,4\right\}$ (marked in green, on the left side of the network) and of cardinality $4$ on nodes $S_2=\left\{8,9,10,11\right\}$ (marked in red, on the right side of the network). 
    
    The \textit{clique} degree for them is equal to $5$ and $3$, respectively. This is easily calculated by enumerating the number of nodes that are adjacent to at least one node in the clique. Specifically, we have $N(S_1)=\left\{1,5,6,7,8\right\}$ with a cardinality of $5$ and $N(S_2)=\left\{3,12,13\right\}$ with a cardinality of $3$.

    The \textit{clique} closeness for them is equal to $\frac{47}{6}$ and $\frac{13}{2}$, respectively. The calculation is less obvious, but still easy to follow. Specifically, $S_1$ has 5 nodes at a distance of $1$, 4 nodes at a distance of $2$, 1 node at a distance of $3$, and 2 nodes at a distance of $4$, for a total calculation of $5\cdot \frac{1}{1}+4\cdot \frac{1}{2}+1\cdot \frac{1}{3}+2\cdot \frac{1}{4}$. Similarly, $S_2$ has 3 neighboring nodes at a distance of $1$, 5 nodes at a distance of $2$, and 3 nodes at a distance of $3$ for a total of $3\cdot \frac{1}{1}+5\cdot \frac{1}{2}+3\cdot \frac{1}{3}$. 
    
    Finally, the \textit{clique} betweenness for them is equal to {$38$ and $31$, respectively}. For this calculation, we first need all shortest paths between any two nodes outside $S_1$ (and later outside $S_2$). {By construction of the graph,} there are 11 shortest paths from any node to all other nodes: that is, a shortest path to any node except for the three nodes in $S_1$. For nodes $1, 5, 6, 7$ all of their shortest paths will use at least one node in $S$. On the other hand, the shortest paths originating from nodes $8,9,10,11,12,13,14,15$ do not use any node in $S$, unless the destination is one of nodes $1,5,6,7${, which have already been accounted for}. {Using these, we have a total betweenness score of $11+10+9+8=38$.}
    

    Similar calculations lead to the betweenness of $S_2$. Specifically, node $12$ will always use at least one node in $S_2$; nodes $13,14,15$ will use at least one node in $S_2$ when the path terminates at any node in $\left\{1,2,3,4,5,6,7,12\right\}$; finally, nodes $1,2,3,4,5,6,7$ will use at least one node in $S_2$ for their shortest paths {only} towards nodes $12,13,14,15$. {Using these, we have a final score of $4+4+4+4+4+4+4+3+31$.} 

    {Once again, a normalized value can be calculated by dividing by the total number of pairs considered, which would be $66$ and $55$ for $S_1, S_2$, respectively. Hence, the normalized values would be equal to $38/66=0.576$ and $31/55=0.564$}
    
\end{myexample}

In the example of Figure \ref{clique_centrality_example}, we also note that it is not necessary that the structure identified is maximal. A subgraph $G[S]$ is maximal with regards to some property (e.g., inducing a clique as in the example) if it cannot be made larger in size while satisfying that property. For example, nodes $S^\prime=\left\{8,9,10\right\}$ are a clique of size $3$, but they are not forming a maximal clique, as adding node $11$ maintains a clique structure and is of larger cardinality. 

\begin{myexample}
    \label{ex:clique2}
    Note a clique of size $3$ containing nodes $\left\{8,9,10\right\}$. It is not maximal, as it is a subset of the clique $S_2$ from Example \ref{ex:clique1}. That said, it can (despite it not being maximal) have higher centrality than the maximal clique it belongs to. Specifically, in this example, its degree is $4$ (its open neighborhood consists of nodes $3, 11, 12, 13$), which is larger than the degree of $S_2$. Additionally, it could have different closeness and betweenness values. 
\end{myexample}

Example \ref{ex:clique2} reveals the necessity for smartly selecting nodes to add in the structure we are building, as bigger in cardinality does not necessarily imply an improvement in its centrality. Using the definitions from \citep{rasti2022novel}, we can assign a structure centrality value to a node. We say that the structure centrality of a node is the maximum value of the centrality of the structure the node belongs to among all structures it can be part of. For example, node $8$ in Figure \ref{clique_centrality_example} can be part of many cliques: the singleton clique $\left\{2\right\}$, the ``edge'' cliques $\left\{3,8\right\}, \left\{8, 9\right\}, \left\{8, 10\right\}, \left\{8, 11\right\}, \left\{8, 12\right\}$, the ``triplet'' cliques $, \left\{8, 9, 10\right\}, \left\{8, 9, 11\right\}, \left\{8, 10, 11\right\}$, and the maximum clique $S_2=\left\{8,9,10,11\right\}$. The clique that leads to the maximum centrality value is the clique centrality of that node. For more details, see the calculations and definitions for other structures (e.g., stars, representatives, quasi-cliques) and centrality metrics in {Rasti and Vogiatzis} \citep{rasti2022novel}.


\section{Brief history on group centrality}
\label{History}
\begin{figure}[!htp]
    \centering
    \includegraphics[width=0.97\textwidth]{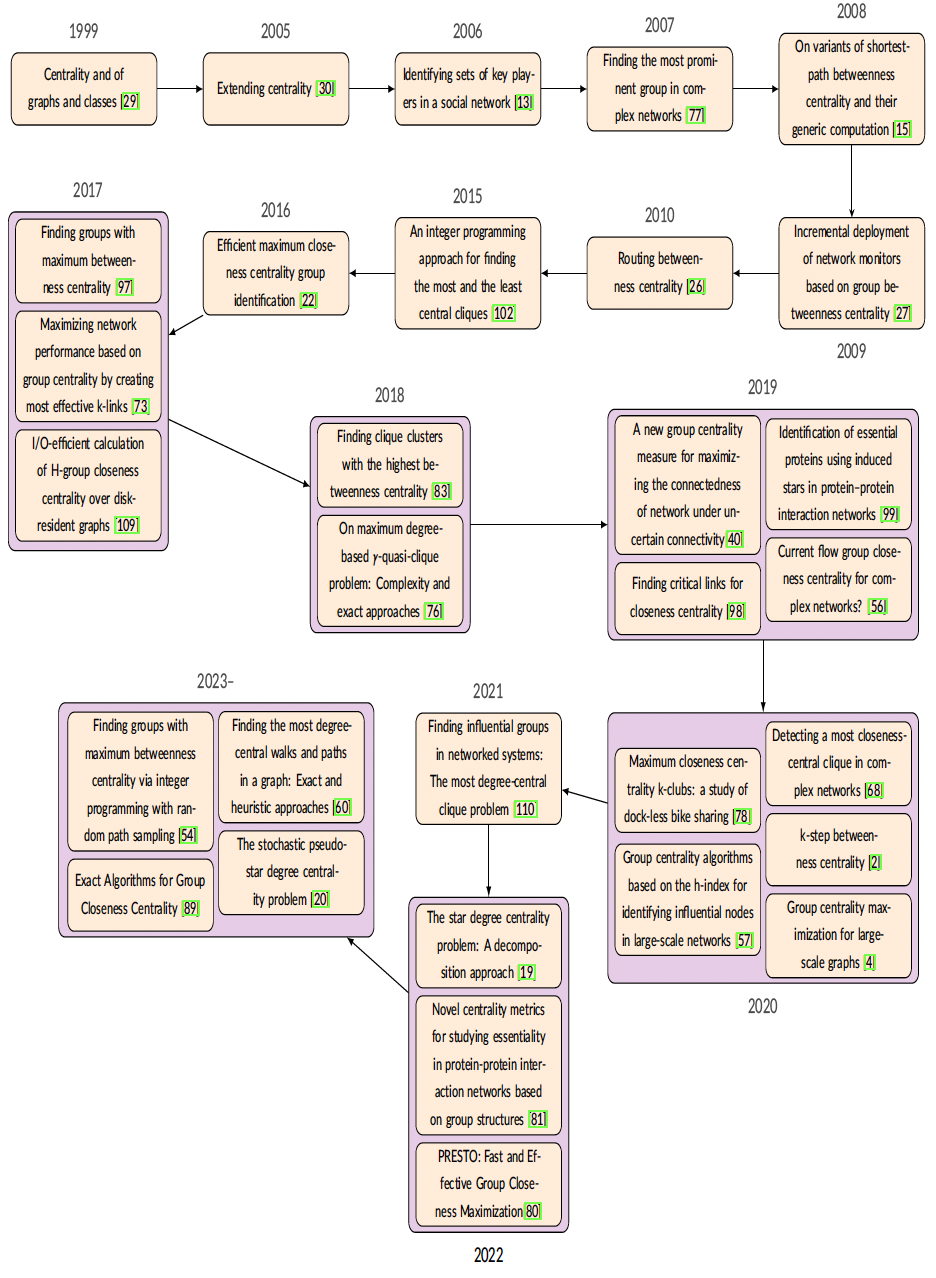}
    \caption{Emergence of OR scientists in group centrality metrics studies over the last two decades.}
    \label{evolution_group_centrality}
\end{figure}

In this section, we provide a review of a subset of studies related to group centrality metrics. Our goal is to briefly describe how they are evolved and are adapted by OR scientists over time. To start with, Figure \ref{evolution_group_centrality} illustrates a chronological overview of group centrality metrics including optimization techniques. We acknowledge that this figure may not capture all relevant studies on group centrality metrics. Indeed, we do mention some works later that were omitted from the figure and explain our reasoning. As mentioned previously, our primary goal is to predominantly focus on OR studies whose number has been gradually increasing in recent years.

{Our criteria for inclusion are as follows: the work must focus on an established or proposed centrality metric, extended and applied to a group of nodes, and solved using common OR techniques. For example, the work by Rysz et al.~\cite{rysz2018finding} satisfies all three criteria: the focus is on betweenness centrality, extended to groups of nodes forming cliques, and the solution technique proposed is a traditional OR technique for mathematical programs in the form of combinatorial branch-and-bound. On the other hand, the fantastic contribution by Medya et al.~\cite{medya2018group} is not included in our survey, despite its use of group betweenness as an objective function and its solution techniques being in the realm of OR. The reason is that the primary focus of the work is not on group centrality, but instead on network design. Specifically, their goal is to improve on a group centrality metric through the careful addition of edges in the network.}

\subsection{{Foundational Studies in Group Centrality}}
Pioneering the field are Everett and Borgatti \citep{everett1999centrality} who are the first in the OR literature to discuss extending centrality and importance to groups of nodes in a network. They motivate the question for groups that are known a priori. We quote from their work: \textit{``Are the lawyers more central than the accountants in a given organisation's social network? Is one particular ethnic minority more integrated into the community than another? To what extent are particular groups or classes (women, the elderly, African-Americans, etc.) marginalized in different networks?''}. This a priori availability of group information that can help us derive descriptive information about the group itself. However, as the authors also note in that same article, we may also focus on identifying groups based on the structure of the network itself \citep{everett1999centrality}. They do mention cliques as a motivation for clique centrality; see the later work by Vogiatzis et al.~\citep{vogiatzis2015integer}.

In their original \citep{everett1999centrality} and subsequent work \citep{everett2005extending}, they base their proposed metrics to nodal centrality metrics. Specifically, they extend traditional metrics in the form of degree, closeness, betweenness, and flow centrality to accommodate a group setting. The authors and subsequent studies have shown how individual nodal calculations for centrality do not necessarily capture  calculations that are necessary for group centrality. In the early 2000s, a fundamental study by Newman and Girvan \citep{newman2004finding} (not included in Figure~\ref{evolution_group_centrality}) uses betweenness centrality to identify groups within a network. They do so by calculating edge betweenness and removing the edge of highest score until the network is disconnected. Importantly, this metric is recalculated after {every} removal. As this work has an arguably different focus than our review, we do not include it in our history. For completeness, though, and as readers may want to investigate the intersection between \textit{community detection} (see also \ref{subsec:excluded}) and centrality, we offer the interesting comparison made by Danon et al. in \citep{danon2005comparing}.

Continuing with these earlier years of group importance,  Borgatti \citep{borgatti2006identifying} utilizes group centrality in social network analysis to solve a \textit{key player} problem. The key player problem (also known as the critical node detection problem) tasks itself with identifying a set of nodes that are ``critical'' for the well-being or success of a network. More recent extensions have since focused on identifying sets of network elements (nodes, edges, cliques). {Borgatti}~\cite{borgatti2006identifying} proposes one of the very first combinatorial optimization approaches in this domain. To our best knowledge, this is one of the earliest studies where optimization algorithms (in the form of genetic algorithms) are employed in the context of group centrality. The importance of players are measured through the impact of their absence on the graph cohesiveness. The author also discusses how degree, betweenness, and closeness metrics may fall short when applied ``as is'' for this problem definition of key players.  Subsequently, in \cite{puzis2007finding}, {Puzis et al.} study the problem of determining groups with the largest  betweenness in a large-scale network. They apply their algorithms on a network service provider infrastructure data set to optimally distribute  intrusion detection devices.

Another prominent work by Brandes \citep{brandes2008variants} presents a wide range of variants of betweenness centrality including proximal betweenness, bounded-distance betweenness, distance-scaled betweenness, and edge betweenness. That same work also presents an algorithm to compute the  group betweenness of a ``given" group. It is critical to mention that the paper focuses on computing the centrality of a fixed group. This is an important step, and one that subsequent studies have used. In this work, the high complexity associated with identifying the optimal group in terms of group betweenness is also mentioned as an important factor. Thus, the study lacks algorithmic details in detecting the most betweenness group, different than {Puzis et al.}~\cite{puzis2007finding}. 

Nonetheless, {Dolev et al.}~\cite{dolev2009incremental}  further this research by studying group betweenness centrality measured as the information flow in an evolving network. The authors propose a heuristic approach with an  approximation guarantee that generalizes the heuristic algorithm proposed in {Puzis et al.} 
 \cite{puzis2007finding} for a similar problem setting. Later, {Fink and Spoerhase} 
 \cite{fink2011maximum} improve the analysis of {Dolev et al.}~\cite{dolev2009incremental}. However, since none of these algorithms scale well in complex networks, in \cite{mahmoody2016scalable}, the authors design a randomized algorithm guaranteeing an approximation of $1-1 /e - \epsilon$ where $\epsilon > 0$. In the following year, {Dolev et al.}~\citep{dolev2010routing} again extend the notion of group betweenness centrality to routing betweenness centrality for a traffic control network. 
Notably, up to this point in our history, there have been no studies to focus on specific structures like cliques or walks (see Section \ref{Definitions}), nor do they propose optimization models.

\subsection{{Early Advancements and Optimization Models}}
{As the field matured, subsequent studies increasingly focused on more sophisticated group centrality metrics, incorporating combinatorial optimization techniques.} One of the first OR studies in this domain is conducted and presented {by Vogiatzis et al.}~\cite{vogiatzis2015integer}. Specifically, the study investigates the most and least important cliques in networks based on three group centrality metrics introduced in Section \ref{Definitions}: degree, closeness, betweenness. The innovation of the work lies in the development of integer linear programming models for each variant in a compact manner. To achieve that, the authors impose an extra requirement of the size of the cliques that are being studied: this requirement has since been dropped from more recent works. Additionally, contrary to much of the more recent literature, the authors propose three variants for betweenness centrality, viewed as optimistic, pessimistic, and probabilistic. Assume that there exist more than one geodesic paths connecting two nodes, and at least one of them uses some node in the selected clique. Then, the optimistic view of betweenness  assumes that this is the only shortest path used; the pessimistic view of betweenness  assumes that this is not the shortest path used and another one, which is not using a node in the clique is used; finally, the probabilistic version is the traditional view of betweenness.

Previous studies until 2016 primarily focus on group betweenness (with the exception of clique centrality in 2015). At this point though, the focus switches towards closeness centrality. First, {Chen et al.}~\cite{chen2016efficient} examine group closeness centrality in order to identify a set of $k$ nodes with the highest group closeness score. The authors first propose a greedy heuristic with an approximation guarantee. To handle large-scale networks, they then develop an order-based sampling algorithm. Thereafter, we observe an increase in OR research in group centrality. We note that {Bergamini et al.}~\cite{bergamini2018scaling} later improve the greedy heuristic designed by {Chen et al.} 
 \cite{chen2016efficient} for the same problem through developing new techniques (bit-level parallelism, submodularity improvement). We prefer not to include this study in Figure \ref{evolution_group_centrality}. It is important to mention that group closeness is highly related to another important OR topic, that of facility location in the form of the $p$-median problem \citep{hakimi1965optimum} {(see subsection \ref{subsec:excluded})}. 

{Veremyev et al.}~\cite{veremyev2017finding} study group betweenness centrality and a number of variants including the length-scaled and the bounded-distance betweenness centralities. The authors examine both group of nodes and cohesive groups ($k$-clubs) and propose mixed-integer linear programming models to solve the problem.  Next, Zhao et al.~\citep{zhao2017efficient} introduce H-group closeness centrality and investigate how to maximize this metric in large disk-resident graphs. This group centrality metric aims to identify a fixed sized node group with  a distance restriction of H, which is the combination of both group degree and group closeness concepts.

Still in 2017, {Ohara et al.}~\cite{ohara2017maximizing} address a group closeness centrality concept. 
The application studied in the manuscript ties to  evacuation performance in a given network. Specifically, the authors are interested in identifying which links to try to design (install) in order to maximize the evacuation performance. In their work, the authors show how their problem can be reduced to a node selection problem. It is the node selection phase that is performed through group centrality computation via a heuristic algorithm with the same approximation guarantee as will be shown in later studies (see, e.g., \cite{vogiatzis2019identification}). In fact, this resemblance in approximation guarantees from the greedy algorithm comes straight from submodularity and leads to an approximation guarantee of 1 - $\frac{1}{e}$.

\subsection{{Emergence of Optimization Algorithms}}
{The evolution of group centrality has been marked by significant advancements in optimization models and algorithmic strategies.}
As an extension to the advancements in {Vogiatzis et al.}~\cite{vogiatzis2015integer} and  {Veremyev et al.}~\cite{veremyev2017finding}, the study conducted by {Rysz et al.}~\cite{rysz2018finding} focuses on determining the most betweenness central clique in a connected network. The important contribution here is that, in contrast to previous works where the clique size is predefined by the authors, the cardinality restriction is lifted. As is common in the literature up to now, they first develop an integer programming model. At a next step, to speed up the optimization phase, the authors then design and employ a combinatorial branch-and-bound (CBB) method. Indeed, following their work, we see a significant uptick in the number of studies utilizing CBB in group centrality literature. This fact shows that smart construction approaches can have quite the place in the group centrality literature.

Interest in clique and clique relaxation problems and their intersection with centrality only continues to grow.  {Pastukhov et al.}~\cite{pastukhov2018maximum} study the problem of identifying a $\gamma$-quasi-clique with the highest degree. The authors first present a linear integer programming model, and then devise two exact solution approaches: a) a CBB algorithm, and  b) a degree
decomposition algorithm.

So far, we are observing a smaller number of OR researchers  working with group centrality with a focus on optimization. This is about to change in 2019, when we witness the first surge in novel problem definitions within the OR community.  {Li et al.}~\cite{li2019current} expand the concept of flow closeness centrality, proposed by {Brandes and  Fleischer}  \cite{brandes2005centrality}, for groups of nodes. This study carries extra significance as it includes current flow in paths other than the shortest paths in the calculation of closeness. Theoretically, the authors also derive the complexity of their problem, and propose two approximation algorithms based on the greedy algorithm. {We also refer the reader to {Bubboloni and Gori}~\cite{bubboloni2022paths} where the authors utilize the concepts of paths and flows to introduce new group centrality measures. However, they do not provide an optimization model or solution approaches, nor do they reference any work presented in Figure~\ref{evolution_group_centrality} that was conducted after 2006.  }

Continuing with closeness, and similar to {Ohara et al.}~\cite{ohara2017maximizing},  {Veremyev et al.}~\cite{veremyev2019finding} explore a new problem in edge detection and closeness centrality. Specifically, they focus on the criticality measured by the influence of an edge's removal on the closeness centrality of nodes connected through these edges. While {Ohara et al.}~\cite{ohara2017maximizing} concentrates on new edge additions to the network, {Veremyev et al.}~\cite{veremyev2019finding} study the implications of edge removals  in relation to group closeness centrality. We note that both node and edge removal, in terms of network element criticality, are broad research topics that we refrain from delving into more in our study. Interested readers are pointed to the immense literature of network element criticality and its applications (see, e.g., \cite{walteros2012selected,lalou2018critical}). {We mention these more concisely in subsection \ref{subsec:excluded}.}

\subsection{{Trends and Novel Group Centrality Metrics}}
{Emerging trends in group centrality research have introduced novel concepts and metrics, such as probabilistic models and new structural definitions, reflecting the continuous innovation within the field.}
Following up, {Vogiatzis and Camur}~\cite{vogiatzis2019identification} propose a novel group centrality metric named Star Degree Centrality (SDC) where the goal is to identify the induced star with the maximum degree in a given network. The authors propose an optimization model and two approximation algorithms to demonstrate that SDC can effectively detect essential proteins in protein-protein interaction networks. The significance of this work  is twofold. Firstly, it is one of the first OR studies applied in a large-scale, real-world network (motivated by protein-protein interaction networks). Their extensive computational results reveal how group centrality can play quite an important role in real-life networks, especially in contexts where nodal metrics fail. Secondly, it investigates a new cohesive group (i.e., star) following the study of cliques and their variants. Stars have a vital/critical element in their center, which makes them very unique structures in a number of applications. 
In  a different, concurrent work, {Fushimi et al.}~\cite{fushimi2019new} propose connectedness centrality under stochastic edge failure, which bears similarities to group closeness centrality. The authors apply this new metric to road networks.

Continuing with 2020, {Taleqani et al.}~\cite{rahim2020maximum}  focus on maximizing closeness centrality for a connected group of nodes with a restricted diameter, also called a $k$-club (see Section \ref{Definitions}), a similar setting observed in {Veremyev et al.}~\cite{veremyev2017finding} for betweenness. A novelty is in the application: dockless bike-sharing systems allow users to ``park'' their bikes anywhere, which makes the $k$-club of highest closeness particularly desirable for both users (``close'' to every node in the network) and bike-sharing providers (restricted diameter to collect and find bikes). This study stands out as the first to apply closeness centrality to $k$-clubs. As a side note, there is an extensive research (see, e.g., \cite{shahinpour2013algorithms}) on identifying maximum cardinality of a $k$-club (which, recall, is a clique relaxation and reduces to the clique problem for $k=1$). 

Within the same year, {Angriman et al.}~\cite{angriman2020group} propose a new group centrality called GED-Walk centrality. This metric is motivated by Katz centrality \citep{katz1953new}: as such, it focuses on walks of all lengths instead of solely the shortest paths, offering an alternative to closeness and betweenness centrality measures.  Following this, {Akgün and Tural}~\cite{akgun2020k} introduce an extension of group betweenness centrality called   $k$-step centrality, where shortest path lengths are restricted by $k$. It is important to note that {Zhao et al.}~\cite{zhao2017efficient} also place length restrictions on shortest paths in their study of group closeness centrality, yet they are not cited by {Akgün and Tural}~\cite{akgun2020k}. The methodology used by {Akgün and Tural}~\cite{akgun2020k} incorporate algorithms developed by {Brandes}~\cite{brandes2008variants} and {Puzis et al.}~\cite{puzis2007finding}. In another study, {Li et al.} \ \cite{li2020group} propose a new group centrality metric named h-index  group centrality, which incorporate the concept of h-index (see \cite{hirsch2005index}) into groups.

Still discussing works in 2020, we also find another group centrality study for cliques. {Nasirian et al.}~\cite{nasirian2020detecting} expand the most-closeness clique problem proposed by {Vogiatzis et al.}~\cite{vogiatzis2015integer} into the concepts of maximum-distance-closeness centrality and  total-distance closeness centrality. They reference {Zhao et al.}~\cite{zhao2017efficient} with the claim, \textit{``The maximum H-group closeness centrality problem is to detect a set of vertices of size at most H with the maximum group closeness centrality in the network"}.
However, we do not think that this claim is accurate since H-group closeness centrality is concerned with a combination of group degree and group closeness centrality with H being a constraint on the lengths of the shortest paths.  The study conducted by {Nasirian et al.}~\cite{nasirian2020detecting} shows similarity to {Rysz et al.}~\cite{rysz2018finding}, as they also remove the restriction of a fixed-sized clique, but instead focus on the most closeness-central clique, as opposed to the most betweenness-central clique.

As discussed earlier, {Vogiatzis et al.}~\cite{vogiatzis2015integer} propose three variants of most influential cliques with respect to degree, closeness, and betweenness. While clique betweenness is studied by both {Veremyev et al.}~\cite{veremyev2017finding} and  {Rysz et al.}~\cite{rysz2018finding}, the topic of clique closeness is covered by  {Nasirian et al.}~\cite{nasirian2020detecting}. The missing piece is completed by {Zhong et al.}~\cite{zhong2021finding} where the authors study the most degree-central clique problem and relax the constraint on the maximum allowed size of the clique.

\subsection{{Recent Developments in Group Centrality}}
{Recent years have witnessed significant advancements in the field of group centrality, characterized by the introduction of innovative metrics and refined optimization methods.}
In 2022, {Camur et al.}~\cite{camur2022star} propose a more effective formulation along with an exact solution technique (i.e., Benders Decomposition \citep{rahmaniani2017benders}) for the SDC problem, proposed by {Vogiatzis and Camur}~\cite{vogiatzis2019identification}. They also present types of networks where the SDC problem is solvable in polynomial time. Additionally, {Rasti and Vogiatzis}  \cite{rasti2022novel} investigate stars, cliques, and representative structures with respect to the same three group centrality metrics. The authors further define a representative set as a group of nodes centered at a node; {i.e.,} it can be viewed as a relaxation of the star, allowing one node to serve as a center while the leaves may or may not share edges between each other. They  propose CBB algorithms for all three structures and all three metrics. Furthermore, {Rajbhandari et al.} 
 \cite{rajbhandari2022presto} revisit the group closeness maximization problem for directed graphs that do not require connectedness, proposing a heuristic algorithm. 
 
Unsurprisingly, in 2023, we witness even more studies in group centrality research. To start with,  {Camur et al.}~\cite{camur2023stochastic} integrate probability values regarding edge connectivity and introduce the very first ``probabilistic" group centrality metric as stochastic pseudo-star degree centrality. According to the authors a pseudo-star is a set of nodes forming a star with at least a probability $\theta\in\left[0,1\right]$. This probability can be viewed as the equivalent of the probability of all edges between the designated center and stars to exist, as well as no edges between any two leaves existing. We note here the similarity between the pseudo-star and a probabilistic clique \citep{miao2014exact}. A probabilistic clique is defined as a set of nodes that form a clique with at least probability $\theta\in\left[0,1\right]$. 

Returning to the pseudo-star degree centrality problem \citep{camur2023stochastic}, this novel group centrality metric is shown to be more effective than its deterministic correspondent (i.e., SDC, see \cite{vogiatzis2019identification}) in determining essential proteins in protein-protein interaction networks. That said,  this study does not employ any stochastic optimization modeling or techniques. On the contrary, the authors utilize a priori available probability values (obtained in the context of the practical application from protein interaction frequencies) to identify the highest weighted degree pseudo-star. 

Then,  {Matsypura et al.} 
 \cite{matsypura2023finding} propose other novel group centrality metrics called most degree-central walks and paths. The study indicates that while the most central shortest paths can be identified in polynomial time, most central walks (e.g., paths) with a pre-defined length remains $\mathcal{NP}$-complete.  It is important to notice that representative sets of nodes, as introduced in {Rasti and Vogiatzis} 
 \cite{rasti2022novel}, do not cover walks and paths. The authors impose a specific connectivity constraints among the nodes  in the subgraph. Thus, this can be classified as a distinct motif different from a clique and a star (see Figure \ref{fig:structures}(a) and (b), compared to (c) and (d), respectively). 

After {Chen et al.} 
 \cite{chen2016efficient} introduce group closeness centrality from an optimization perspective, {Staus et al.}  \cite{staus2023exact} propos two exact solution techniques for the same problem. Lastly, {Lagos et al.}~\cite{lagos2024finding} revisit the problem of identifying groups with maximum betweenness centrality as previously discussed by {Puzis et al.}~\cite{puzis2007finding} and {Veremyev et al.}~\cite{veremyev2017finding}. The authors integrate mathematical modeling with randomized path sampling.  


\subsection{{Excluded Aspects of Group Centrality}} 
\label{subsec:excluded}

{Throughout the previous subsections, we have mentioned a series of works that are relevant and related, but were excluded from this survey. These included the detection of critical nodes and sets of graph elements, graph partitioning and community detection alongside ``cut'' problems, facility location problems, and coverage problems on graphs. We provide a list of indicative works and surveys in the next paragraphs to guide researchers that are interested in these areas and their relationships to group centrality.}

{Critical nodes and graph elements are very related to our survey and focus, which is also the reason why Figure \ref{evolution_group_centrality} does include the foundational work ``\textit{Identifying sets of key players in a social network}'' by Borgatti in 2006 \cite{borgatti2006identifying}. However, our focus remains firmly on the realm of centrality (which is usually a proxy for importance or criticality), hence we do not fully discuss a series of works in this area. Interested readers will find the surveys by Walteros and Pardalos in 2012 \cite{walteros2012selected} and by Lalou et al. in 2018 \cite{lalou2018critical} particularly useful.}

{Another part of the related literature includes graph communities. Communities are often graph structures that are of interest in our survey, e.g., in the form of cliques or quasi-cliques. Earlier in this work, we mentioned how betweenness centrality served for the detection of community structures in networks \cite{newman2004finding}. Newman is also responsible for one of the earliest comprehensive surveys of identifying communities in networks \cite{newman2004detecting}. A very recent and thorough survey on a variety of tools and applications of community detection is by Fortunato \cite{FORTUNATO201075}. While centrality is often used as a way to identify nodes/edges to consider while detecting communities (see, e.g., Girvan-Newman method and betweenness \cite{newman2004finding} or Fortunato method and current flow centrality \cite{fortunato2004method}), the focus of these studies is on community detection rather than centrality, so we exclude them from discussion. Sometimes, especially in problems involving community detection through cuts (see, e.g., normalized cuts \cite{shi2000normalized} and other variants), the focus can be on communities of interest to this survey (e.g., the structural requirements in \cite{vogiatzis2018integer}), but there is no mention of centrality, so these are also excluded from our discussion.}

{Additionally, location and siting problems have long benefited from the use of centrality metrics. Specifically, Angriman et al.~mention that group closeness (which is one of the focal group centrality metrics in this survey) appear to be a particularly useful objective in facility location problems \cite{angriman2020group,angriman2021group}. That said, the literature in facility location is vast (see, e.g., the recent surveys in \cite{mladenovic2007p,daskin2015p}), so we omit these works from our study unless they focus specifically on group centrality metrics with an application to facility location and siting.}

{Finally, we discuss the relationship of group degree to maximum coverage problems in networks.  For example, the study titled ``\textit{Group Degree Centrality and Centralization in Networks}" by {Krnc and Škrekovski}~\cite{krnc2020group} examines the set of nodes with the largest degree. However, the focus on largest-degree subgraphs without any connectivity or structural restrictions aligns more with the maximum coverage problem (see {Megiddo et al.}~\cite{megiddo1983maximum}, which is beyond our current scope. Similarly, the concept of group coverage centrality, mostly studied by researchers outside the field of OR and out of the scope the work, are not included in our review \citep{yoshida2014almost, medya2018group, d2019coverage, komusiewicz2023group}, despite their importance. }

{Before moving into the next section where we continue our discussion with analyzing group structures and optimization methods utilized,  we would like to mention one more aspect of group centrality. During our review, we encountered a group of work in which group centrality is addressed using the Shapley value \citep{hart1989shapley}. This set of work is predominantly considered in social network analysis using game theory, which is another branch of  OR (see \citep{szczepanski2012new, michalak2013efficient, flores2016assessment,sun2020game, belik2023measuring} in order). However, these research products seem to be more independent of the framework that we have presented above. While all of these studies reference {Everett and Borgatti}~\cite{everett1999centrality} regarding group centrality, we did not find any reference to the rest of the works presented in Figure~\ref{evolution_group_centrality}. Thus, we prefer not to include these studies during our analysis in the next section. }



\section{Structures and computational approaches} \label{Solution Approaches}

To begin with, we first discuss the graph structures examined by researchers. We categorize these structures into five groups: group, clique, star, representative set, and walk, which are defined in detail in Section \ref{Definitions}. Regarding group centrality, we report three main types: degree, closeness, and betweenness, along with a variant defined as any centrality measure not related to these three. Table~\ref{Table1} summarizes the studies conducted on each group structure in terms of the aforementioned group centrality metrics.

\begin{table}[ht]
\caption{Comparison of different group structures analyzed for group centrality metrics\label{Table1}}
\centering
\begin{tabularx}{\textwidth}{@{} 
     >{\bfseries}p{\widthof{\textbf{Representative}}} 
     X 
     X 
     X
     X
     @{}}
\toprule
 {} &  \textbf{{Degree}} &  \textbf{{Closeness}} & \textbf{ {Betweenness} }& \textbf{ {Variant}} \\ 
\midrule
\rowcolor{lightgray}
{Group} &  &  \cite{chen2016efficient, ohara2017maximizing, zhao2017efficient, bergamini2018scaling, veremyev2019finding, li2019current,angriman2019local,  angriman2021group, rajbhandari2022presto, staus2023exact}  &  \cite{puzis2007finding, brandes2008variants, dolev2009incremental, dolev2010routing, fink2011maximum, mahmoody2016scalable, veremyev2017finding, akgun2020k, lagos2024finding}&  \cite{borgatti2006identifying, fushimi2019new, angriman2020group, li2020group}  \\ 
\addlinespace
{Clique}  &\cite{vogiatzis2015integer, pastukhov2018maximum, zhong2021finding} & \cite{vogiatzis2015integer,nasirian2020detecting}   & \cite{vogiatzis2015integer,veremyev2017finding, rysz2018finding} & \\ 
\addlinespace
\rowcolor{lightgray}
{Star} & \cite{vogiatzis2019identification, camur2022star, camur2023stochastic}  & \cite{rasti2022novel}   & \cite{rasti2022novel}  & \\ 
\addlinespace
Representative \newline Set & \cite{rasti2022novel} & \cite{rasti2022novel}   & \cite{rasti2022novel} &   \\ 
\addlinespace
\rowcolor{lightgray}
\textbf{Walk} & \cite{matsypura2023finding} &    &  & \\ 
\bottomrule
\end{tabularx}

\end{table}

As previously mentioned, groups with the largest degree centrality are more closely related to the maximum coverage problem literature; thus, the corresponding section in Table~\ref{Table1} does not include any of these studies. Extensive research is conducted on groups of nodes without any restrictions in terms of both closeness and betweenness centrality (see the first row in Table~\ref{Table1}). However, from OR scientists, we observe significant interest in cohesive groups. Specifically, they focus on particular structures, with cliques attracting the most attention. In recent years, the star structure is also  examined from a group centrality perspective. In fact, Rasti and Vogiatzis  \cite{rasti2022novel} generalize all clique and star-type graph structures under representative sets, covering all the major group centrality metrics in that study. Additionally, the authors define group centrality on a nodal level as the value of the group (inducing a specific structure) with maximum centrality value, so long as it includes that node as a member (in the case of cliques) or as a center (in the case of stars). 

Since much of the existing literature covers mostly both star and clique structures, {Matsypura et al.} 
 \cite{matsypura2023finding} examine walk and path centrality, expanding the concept of cohesive groups. This study opens a new window into the group centrality literature from a special structure perspective. Moreover, we note a study by {Grinten et al.} \cite{van2021new}, which introduces a novel centrality measure named Forest Closeness Centrality for a group of nodes. This concept builds upon the Forest Distance Closeness initially proposed by {Jin et al.} \cite{jin2019forest}. However, due to the absence of a formal journal publication on this new measure and its still-developing stage of research, we avoid providing more details about it in our study. Yet, we mention it in the hopes that it will spark an interest in further investigation in very novel structures and centrality metrics.

\begin{table}[!ht]
\centering
\caption{Comparison of different optimization techniques employed for group centrality metrics}
\begin{tabularx}{\textwidth}{@{} 
     >{\bfseries}p{\widthof{\textbf{Branch \& Bound}}} 
     X 
     X 
     X
     X
     @{}}
\toprule

 {} &  \textbf{{Degree}} &  \textbf{{Closeness}} & \textbf{ {Betweenness} }& \textbf{ {Variant}} \\ 
\midrule
\rowcolor{lightgray}
{Mathematical} \newline{Modeling} & \cite{vogiatzis2015integer,pastukhov2018maximum,vogiatzis2019identification,zhong2021finding,camur2022star,  rasti2022novel,matsypura2023finding, camur2023stochastic} & \cite{vogiatzis2015integer,bergamini2018scaling, veremyev2019finding,rahim2020maximum,nasirian2020detecting,rasti2022novel,staus2023exact} & \cite{vogiatzis2015integer,veremyev2017finding,rysz2018finding,rasti2022novel} & \\ 
\addlinespace
{Heuristic} \newline {Approaches} & \cite{fushimi2019new,matsypura2023finding} & \cite{chen2016efficient,veremyev2019finding,rahim2020maximum, rajbhandari2022presto} & \cite{puzis2007finding,dolev2010routing,akgun2020k} & \cite{borgatti2006identifying,fushimi2019new} \\ 
\addlinespace
\rowcolor{lightgray}
{Approximation} \newline {Algorithms} & \cite{vogiatzis2019identification} & \cite{chen2016efficient,ohara2017maximizing,zhao2017efficient,bergamini2018scaling, li2019current, angriman2019local, angriman2021group} & \cite{dolev2009incremental, fink2011maximum, mahmoody2016scalable} & \cite{angriman2020group} \\ 
\addlinespace
{Branch \& Bound} \newline {Approaches} & \cite{vogiatzis2015integer,pastukhov2018maximum, zhong2021finding,rasti2022novel} & \cite{nasirian2020detecting,rasti2022novel,staus2023exact} & \cite{rysz2018finding, rasti2022novel} & \\ 
\addlinespace
\rowcolor{lightgray}
{Decomposition}\newline {Algorithms} & \cite{pastukhov2018maximum,camur2022star}, \cite{camur2023stochastic} & & & \\ 
\bottomrule
\end{tabularx}

\label{Table2}
\end{table}

{
Next, we examine the optimization methodologies adopted by OR researchers. We establish a  distinction between exact and heuristic algorithms to reflect their different roles and efficacy in solving complex problems. Exact methods systematically explore the solution space to guarantee the identification of optimal solutions. During our review, we encountered three main groups: a) mathematical modeling techniques such as integer programming and linear programming, b) branch-and-bound approaches, and c) decomposition algorithms.}

{
Conversely, heuristic methods, which we segment into simple heuristics and approximation algorithms in this study, prioritize computational efficiency and practical applicability over guaranteed optimality. Simple heuristics provide feasible solutions quickly without assurances on closeness to optimality.  Approximation algorithms, as a specialized form of heuristics, offer a balance by providing solutions with known bounds on their deviation from optimality. This categorization aids in understanding the trade-offs between solution quality and computational resources, thereby guiding the selection of appropriate methods based on specific problem requirements and constraints.}

In light of these, we categorize solution methods into five main groups: (a) optimization modeling (e.g., integer programming, linear programming), (b) heuristic algorithms, (c) heuristics with approximation guarantees (i.e., approximation algorithms), (d) combinatorial branch-and-bound (CBB) algorithms (this also includes ``smart selection'' algorithms that use lower and upper bounds), and (e) decomposition algorithms (e.g., Benders Decomposition). For group (a), it is important to note that studies offering a heuristic approach or an exact solution technique are highly likely to include a mathematical model, leading to potential duplication among cells. Regarding centrality, we report the same metrics as shown in Table~\ref{Table1}. As a result, we present Table~\ref{Table2} to categorize recent works based on the methodology adopted.

In Table~\ref{Table2}, we observe that  studies examining clique centrality regardless of the centrality metrics used are predominantly  addressed through mathematical modeling and CBB approaches. First of all, this is due to the fact that cliques (and clique relaxations) are very prominently featured in OR studies and hence, OR practitioners are more prone to employ them. More importantly, though, we note that CBB and other smart selection approaches are well-suited for clique problems due to their hereditary property. Constructing a larger clique by adding a node or by combining multiple smaller cliques can be done with a quick feasibility check, making CBB a good candidate.

Conversely, research related to highest degree star centrality frequently employ decomposition algorithms to solve the large-scale models. In both {Camur et al.}  \cite{camur2022star} and {Camur et al.} \cite{camur2023stochastic}, the authors implement various Benders decomposition strategies, including traditional Benders approach, modern Benders, two-phase, three-phase decomposition, and Logical Benders. The problems of star centrality with the highest closeness and betweenness, on the other hand, are tackled by CBB approaches.

It is evident that studies lacking any specific motif are often tackled using heuristic or approximation algorithms. In essence, there are additional studies regarding groups without motifs that we do not include in Figure~\ref{evolution_group_centrality}. For instance, the concept of group closeness, studied by  {Chen et al.} \cite{chen2016efficient}, is later adapted by other researchers  to  improve the approximation guarantee.   {Bergamini et al.} \cite{bergamini2018scaling} then scale up the existing approximation, and {Angriman et al.} \cite{angriman2019local} offer an algorithm that is  one to two orders of magnitude faster, based on a new local search heuristic. Subsequently, {Angriman et al.} \cite{angriman2021group} provide  a constant approximation for the same local search algorithm. 

We initially aimed to provide another table categorizing the existing research based on application areas. However, we discover that the majority of work on group centrality, especially those incorporating optimization techniques, focuses on randomly generated networks. These include networks models including the \emph{Barab{\'a}si–Albert}, \emph{Erdős–Rényi}, and \emph{Watts–Strogatz} models. Such networks are randomly generated, which allows for controlled experimentation and analysis but may not always accurately reflect real-world scenarios. As we note in our directions for future work (Section \ref{Future Work}), we would like to see more group centrality metrics that are naturally derived from emerging applications and emerging technologies around us.

Nevertheless, there are few exceptions where group centrality metrics are applied to real-world scenarios. Specifically, we observe significant applications in biological networks, particularly in protein-protein interaction networks (refer to \cite{vogiatzis2019identification, camur2022star, rasti2022novel, camur2023stochastic}). These studies utilize group structures such as  star and representative sets in analyzing protein-protein interaction networks to determine essential proteins for the survival of the organisms.

\section{Future work} \label{Future Work}

We observe that the OR community has seen significant advancements on group centrality in the last decade as discussed in Section \ref{History}. In addition, we provide additional analysis to provide more depth insights in Section \ref{Solution Approaches}. Yet, there are several unexplored areas that might of interest to optimization experts. As recently introduced by {Matsypura et al.} \cite{matsypura2023finding}, both walk and path group centrality are evaluated primarily from the largest degree perspective. Thus, there is a research gap for these novel group centrality metrics with respect to closeness and betweenness centrality that we believe needs to be addressed.

Another research direction lies in the domain of tree structures within graphs.
Trees, defined as connected graphs where there exists a unique path between each node pair, as a special type of graph, could serve as a basis for developing new group centrality measures.  While the concept of spanning tree centrality exists in literature, it differs from our understanding in this paper. The existing studies focus on the probability of an edge being used in a uniformly sampled spanning tree \citep{teixeira2013spanning, hayashi2016efficient}. Identifying ``central'' trees could serve in application as developing communication protocols; their relationships to connected dominating sets \citep{guha1998approximation,validi2020optimal} could also be further investigated.

Further, examining motif structures like cliques and stars through modified centrality metrics, such as weighted versions, can be a fruitful direction. Identifying these structures in relation to new centrality metrics is a promising area for future research. For example, researchers might be interested in investigating variants of centrality metrics (e.g., the last column of Tables~\ref{Table1} and~\ref{Table2}) when it comes to cohesive groups. This interest arises because we no study examining these groups beyond the scope of the three major group centrality metrics is encountered during our review.

{Newman} \cite{newman2005measure} introduces a new measure of betweenness centrality based on random walks. This measure considers the contribution of all paths, not just the shortest ones, while placing more weight on the shortest paths and considering generating random walks (see {Noh and Rieger} \cite{noh2002stability}). Later, {Gillani et al.} \cite{gillani2021group} extend this idea to group to group version where centrality of a node is measured through a set of source and a set of sink nodes.   They indicate that ``\textit{We look forward to extend random walk based centrality definition
to the group centrality setting in which we measure the centrality
of groups rather than individual nodes.}". 
We believe that this might be another avenue for further research. Even though this metric might sounds similar to the GED-centrality proposed by {Angriman et al.} \cite{angriman2020group}, we note that random-walk centrality is more related to algebraic-based metrics including the Laplacian equations.

As for the solution techniques, we note that there is no study proposing exact solution approaches for groups without motifs. Investigating this could lead to significant contributions. Additionally, there is a clear path towards exploring probabilistic group centrality metrics, such as examining the problem of maximum probabilistic clique \citep{miao2014exact} from a group centrality perspective.  Assuming static, unchanging conditions in large-scale network can be a significant limitation in real-world scenarios where uncertainty and variability are inherent. That is why exploring this would be important to understand network dynamics better.

Considering that current models in OR literature are predominantly deterministic, introducing stochastic optimization techniques, like two-stage stochastic models, could be a valuable addition to the field.  Such models can help in identifying influential groups that are crucial for the robustness and efficiency of a network under probabilistic conditions. This approach could lead to more resilient and adaptive network designs, thereby enhancing decision-making processes in areas such as infrastructure planning, supply chain and logistics. Additionally, considering the fact that for many networks we do not have a full picture of all the existing nodes and edges (see, e.g., illicit networks which have attracted significant recent interest \cite{kosmas2023transdisciplinary,anzoom2023uncovering}), stochastic extensions to group centrality can lead to more realistic identification of key actors in networks with hidden elements.

In addition,  it is evident that the majority of the research in group centrality and optimization remains within the realm of theoretical research. Thus, we find it important to draw interest in applying these concepts to real-world networks, such as in social network analysis and the identification of influential actors, transportation systems and their resilience, networks encountered in socio-technical systems, or biological networks, to increase visibility and promote interdisciplinary research opportunities. Specifically, we mention a few directions next.

We start from a very important work by {Matsypura et al.} \cite{matsypura2023finding}: identifying paths and walks of high degree can have applications in using autonomous vehicles to monitor a number of targets (by approaching them). As autonomous vehicles are becoming more  prominent in applications such as humanitarian relief operations, last mile deliveries, health visits, etc., we believe that the study of closeness metrics from the perspective of walks, paths, and tours/cycles will be an important piece of the puzzle. 

Nodal centrality has been prominent in interesting new societal problems, such as investigating criminal and illicit networks \citep{bichler2017drug,grassi2019betweenness}. That said, group centrality metrics present a unique opportunity to identify many key illicit actors at once. Moreover, we mentioned earlier the need for more stochastic measures of group centrality. These would be particularly interesting in this context, due to the large number of hidden or obscured entities and relationships.

Additionally, a number of researchers have investigated criticality and centrality from the perspective of disaster management (mostly evacuation). Continuing on that path, we believe it is important to include refueling constraints (see, e.g., {Vogiatzis and Kontou} \cite{vogiatzis2024critical}) in the calculation of group centrality when applied to evacuation, especially considering the scarcity of gas (and soon other alternative fuels) during an emergency. Finally, another avenue for research is in pedagogy and the formation of groups in the classroom (see, for example, {Hill and Peuker} \cite{hill2023expanding}). Utilizing group centrality to identify groups with ``natural leaders'' or groups that satisfy a specific connectivity constraint (e.g., high degrees to other groups or mentors, high closeness to some mentor, etc.) can improve team formation and lead to improved outcomes inside and outside the classroom.

\section{Conclusion} \label{Conclusion}
In this work, we review studies related to group centrality, focusing on identifying the most influential set of nodes in a network using predefined metrics such as degree, closeness, and betweenness. Our emphasis is on approaches that utilize optimization modeling (e.g., {integer} programming) and solution techniques, including heuristics, approximation algorithms, decomposition methods, and combinatorial branch-and-bound methods.

We begin with providing preliminaries and definitions related to centrality and group centrality. We then share  a comprehensive discussion on the adoption of the group centrality concept by OR researchers over time and indicate its rapidly growing interest. We then discuss how group centrality is studied in various structures, including cliques and stars, and summarize the preferred optimization techniques for different group centrality metrics.

Our findings reveal that OR researchers are primarily interested in identifying cohesive groups with connectivity restrictions. However, there is an important lack in real-world applications, as algorithms and models are predominantly tested on randomly generated, synthetic networks. 

We  identify several other research gaps. Firstly, there are few studies focusing on probabilistic group centrality metrics, where the existence of edge connections or nodes may include randomness. Moreover, we have not encountered any studies employing stochastic optimization techniques. It could be interesting to explore how scenario-based stochastic approaches might provide more robust group identification amidst randomness.

\section*{Declaration of interests}
The authors have no competing interests to declare that are relevant to the content of this article.

\bibliographystyle{abbrv_networks}
\bibliography{references_net}

\end{document}